# Global Waterbody Calculator: A high-resolution global database of lakes and reservoirs depth-area-volume (D-A-V) relationships

Shengde Yu[1],*, Yukai Wu[2], Weikun Liao[3], and Philippe Van Cappellen[1,4]

1 Ecohydrology Research Group, Department of Earth and Environmental Sciences, University of Waterloo, Waterloo, ON, Canada

2 The Edward S. Rogers Sr. Department of Electrical and Computer Engineering, University of Toronto, Toronto, ON, Canada

3 Department of Chemical Engineering and Applied Chemistry, University of Toronto, Toronto, ON, Canada

4 Water Institute, University of Waterloo, Waterloo, ON, Canada

* Correspondence: Shengde Yu; Email: s228yu@uwaterloo.ca

## Abstract

The new GLRDAV database assembles key morphometric characteristics for over 1.4 million lakes and reservoirs. By integrating data from HydroLAKES and GLOBathy, > 17 million polynomial (orders 1–5) and power functions describing depth–area–volume (D-A-V) relationships at a 0.1 m depth resolution are presented. The D-A-V relationships are validated against existing databases (ReGeom and GRDL) that provide comparable or simplified global bathymetric relationships, as well as in situ measurements for four waterbodies in the Texas Water Development Board (TWDB). The results show that higher-order polynomial equations (particularly orders 4 and 5) generally yield the lowest RMSE and highest goodness-of-fit (i.e., $R^2$). Although power functions and lower-order polynomials can offer reasonable representations for simpler and shallower systems, they typically underperform when applied to large or morphologically complex lakes and reservoirs. Our findings underscore the flexibility and robustness of GLRDAV, offering a globally consistent, high-resolution, and computationally scalable resource for hydrological, ecological, and water resource management applications.

## 1 Introduction

Global freshwater bodies, including natural lakes and reservoirs, are important components of the Earth's life-support systems. These water bodies are crucial for maintaining ecological and hydrological balances; they modulate connected river flows and the cycling of nutrients and the mitigation of pollutants (Grill et al., 2019; Pi et al., 2022; Yao et al., 2023). The ecosystem health and storage volume of freshwater lakes and reservoirs, however, vary in response to climate gradients, seasonal weather conditions, as well as increasing anthropogenic pressures (Cooley et al., 2021; Klein et al., 2024; Messager et al., 2016). For instance, the effects of human activities, from agricultural irrigation and industrial usage to energy generation for urban development, have caused the aggregated storage volume of the world's largest freshwater bodies (n = 1,972) to shrink by 53% over the period 1992 to 2020 (Wurtsbaugh et al., 2017; Yao et al., 2023). Beyond water scarcity itself, the detrimental consequences include increased greenhouse gas emissions, biodiversity loss, and negative impacts on human health and well-being. These challenges highlight the urgent need for sustainable management and conservation of freshwater resources (Battin et al., 2023; Harrison et al., 2021; Rosa et al., 2020; Shi et al., 2024; Wu et al., 2019).

Models that provide insights into water flow dynamics and water quality of lakes and reservoirs rely on some level of representation of their physical characteristics (Bai et al., 2022; Burigato Costa et al., 2019). Among these, the depth-area-volume (D-A-V) relationships help to capture the essential bathymetric features of a water body (An et al., 2022; Busker et al., 2019; Duan & Bastiaanssen, 2013; Hao et al., 2024) that, in turn, play a key role in the coupled hydrological, biogeochemical, and ecological processes regulating aquatic ecosystem productivity, biodiversity, and resource availability (Khazaei et al., 2022; Li et al., 2019). Hence, when integrated with models, current or future, D-A-V relationships can support a wide range of applications (Hao et al., 2024; Li et al., 2019).

Previous research has used a variety of methodologies for acquiring bathymetric and morphometric data of freshwater bodies (Donchyts et al., 2022; Hao et al., 2024; Lehner et al., 2011; Li et al., 2020; Yigzaw et al., 2018). Each method possesses its own strengths and limitations. One major approach involves imagery-based techniques, such as obtained from satellite observations. The latter has a fundamental limitation, however, because it relies on surface water variations, such as changes in water level, to establish elevation-area (E-A) relationships (Li et al., 2019, 2020; Liu & Song, 2022). Therefore, while satellite imagery techniques can systematically yield data for large numbers of water bodies, they are restricted in their capability of characterizing basin morphology beneath the water surface.

Bathymetric data are still mostly derived from direct mapping of the bottom topography of individual lakes or reservoirs. To enable regional and global scale analyses, attempts have been made to assemble this bathymetric information into searchable databases. For example, Lehner et al. (2011) fitted power function relationships between volume, area, and dam height for 6864 reservoirs when developing their Global Reservoir and Dam Database (GRanD). The initial version of GRanD, however, only provides information on the reservoir storage capacities. More recently, GRanD v1.3, which includes 7,320 reservoirs globally, comprises additional attributes by providing not only information on capacity but also on reservoir surface area and dam construction details (Lehner et al., 2011). Along similar lines, ReGeom offers a storage-area-depth dataset for 6,824 major reservoirs using simplified shapes to approximate their surface areas (Yigzaw et al., 2018).

Expanding beyond reservoirs, the HydroLAKES database has mapped shoreline polygons of 1.42+ million natural lakes and artificial reservoirs to generate surface areas (Messager et al., 2016). HydroLAKES also estimates water residence times and average water depths, although inconsistencies between the different parameters have been pointed out in some cases (Messager et al., 2016). The later GLOBathy significantly advanced the bathymetric information content by providing estimates of maximum water depths, bathymetric maps, and D-A-V relationships for over 1.4 million water bodies (Khazaei et al., 2022). However, the dataset suffers from unit inconsistencies and limited depth resolution in D-A-V relationships. More recently, the GRDL database enhanced the depth resolution of D-A-V relationships to 1 m (Hao et al., 2024). However, GRDL only considers reservoirs, and its use requires significant computational resources.

Here, we present a global dataset of D-A-V relationships for over 1.4 million lakes and reservoirs at a depth resolution of 0.1 m. For each water body, the new GLRDAV dataset offers the choice between 12 equations: six depth-area (D-A) and six depth-volume (D-V) equations. Each set of equations comprises five polynomial (first to fifth order) and one power function. Furthermore, for each 0.1 m depth interval within a given water body, the dataset provides the corresponding coefficients for the equations. To support user-friendly data access, we also developed a Python package that streamlines the retrieval of the 17+ million equations and their coefficients.

## 2 Methods and materials

### 2.1 Foundational databases

The databases used for training and validation purposes are summarized in Table 1. HydroLAKES and GLOBathy are the primary sources for training data (Khazaei et al., 2022; Messager et al., 2016). HydroLAKES employs a geostatistical approach to deliver attributes for 1,427,688 waterbodies, including names, locations, volumes, surface areas, and discharge estimates. GLOBathy builds on HydroLAKES using a GIS-based framework, incorporating location and depth information to estimate bathymetric properties and D-A-V relationships for the same set of waterbodies. These datasets serve as the core inputs for constructing the GLRDAV database. Notably, in the original HydroLAKES dataset, the Caspian Sea stands out as the largest saltwater lake, representing approximately 41% of the total volume recorded in HydroLAKES (Messager et al., 2016).

To validate our results, we employed three independent datasets: ReGeom (Yigzaw et al., 2018), GRDL (Hao et al., 2024), and in situ measurements from the Texas Water Development Board (TWDB; https://www.twdb.texas.gov/surfacewater/surveys/completed/list/index.asp). ReGeom provides D-A-V data for 6,824 reservoirs, while GRDL offers high-resolution D-A-V relationships for 7,250 reservoirs. The TWDB dataset includes observed D-A-V relationships for 121 Texas waterbodies, offering ground-truth reference (Table 1).

**Table 1. Summary of foundational databases for developing GLRDAV.**

| Databases (References) | Number of waterbodies | Main contents | Limitations | Purpose |
|---|---|---|---|---|
| GLOBathy (Khazaei et al., 2022) | 1,427,688 | Name, locations, estimated depth, D-A-V | Wrong units in D-A-V file, simplified depths and D-A-V equations, not applicable for water level fluctuations and other user-defined scenarios | Training and validation |
| HydroLAKES (Messager et al., 2016) | 1,427,688 | Name, locations, volume, surface area, discharge information | Only basic information, no D-A-V | Training and validation |
| ReGeom (Yigzaw et al., 2018) | 6,824 | Name, locations, D-A-V | Simplified shapes of waterbodies, data errors | Validation |
| GRDL (Hao et al., 2024) | 7,250 | Name, locations, D-A-V | Requires high GPU resources for extended training, heavily depend on precision of the original images, only reservoirs, not user-friendly | Validation |
| TWDB | 121 | D-A-V relationships (observations) | Data formats not user-friendly | Validation |

## 2.2 D-A-V relationships and 3D visualization

We integrated GLOBathy and HydroLAKES to construct the training dataset. Regression techniques were employed to fit relationships between depth and surface area (D-A), as well as depth and volume (D-V) across more than 1.4 million waterbodies (Figure 1). To minimize computational overhead and enable seamless coupling with hydrological models, we approximated high-resolution discrete D-A-V data by fitting continuous polynomial and power-law functions. These surrogate expressions allow rapid estimation of waterbody volumes from water levels in modeling applications. To capture nonlinear dependencies, we employed polynomial and power-law regressions. Polynomial equations ranging from first to fifth order were considered, while power-law equations were fitted using logarithmic transformations. For each waterbody, the D-A and D-V properties were captured by a separate set of five polynomial equations and one power function, together with their corresponding coefficients. For each 0.1 m depth interval, the horizontal surface area and volume are also provided. The approach is formalized by the following equations.

$$y = f(Lat, Lon, x, D_{max}, V_{total}) \qquad \text{Equation 1}$$

$$y(x) = \begin{cases} a_0 + a_1 x + a_2 x^2 + \cdots + a_n x^n, \\ \alpha x^{\beta} \end{cases} \qquad \text{Equation 2}$$

where n is an integer from 1 to 5; $y$ is the dependent variable (either surface area or volume), and $x$ represents water depth measured from the bottom up; $\alpha$ and $\beta$ are fitted coefficients of the power function. Additionally, we also provide surface area and volume values at 0.1 m increments of depth. The training process further incorporates each waterbody's geographic coordinates (latitude and longitude, consistent with the HydroLAKES database), maximum depth $D_{max}$ and total volume $V_{total}$, which are extracted from GLOBathy, and HydroLAKES databases.

Python scripts were used to implement the methodology, leveraging Python scientific libraries for efficient data processing and analysis. Pandas were used for handling whole datasets (GLDAV, GLOBathy, and HydroLAKES), Numpy package for the numerical computations, and scikit-learn for the derivation of the regression D-A-V equations. To handle the large number of data, we applied multithreading via concurrent.futures in the ThreadPoolExecutor utility. Interpolation, performed using NumPy, provided the horizontal area and volume estimates for 0.1 m depth interval of a given water body. All scripts, user manuals, and datasets are archived in the Federated Research Data Repository (doi: https://doi.org/10.20383/103.0881).

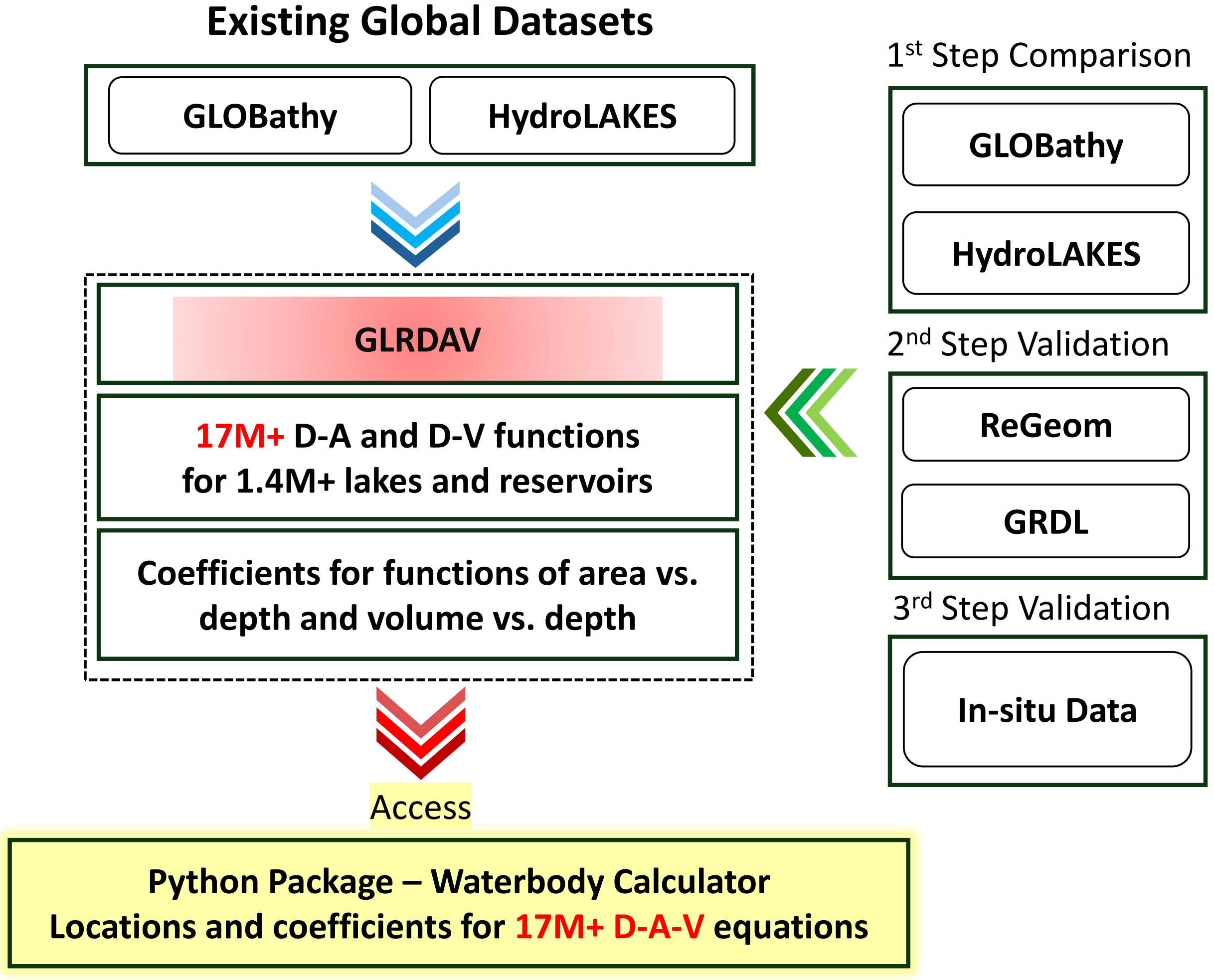

**Figure 1. Conceptual workflow of the GLRDAV framework. Bathymetric and morphometric data from GLOBathy and HydroLAKES are integrated to derive depth–area–volume (D-A-V) equations, and validated against ReGeom, GRDL, and in-situ measurements.**

Bathymetric rasters were produced by remapping calibrated depth–area curves onto a high-resolution grid. Lake polygons from HydroLAKES were first rasterized at 1″ (≈ 30 m) resolution, then each water pixel's distance to the shoreline was computed and used to derive its cumulative area percentile. An interpolator built from the normalized depth–area relationship assigned a depth to each pixel, yielding a continuous bathymetry surface. GeoTIFF outputs were subsequently visualized in three dimensions. Full implementation details and code are provided in the supplementary.

Overall, more than 17 million equations were generated to capture the D-A-V relationships across global lakes and reservoirs. The user accesses the equations via a Python-based tool named "Global Waterbody Calculator," which organizes waterbody identifiers, geographic coordinates, and model coefficients to facilitate the retrieval of the data and equations (Figure 1).

## 2.3 Statistical analysis

To evaluate the accuracy and performance of the derived D–A–V relationships, we employed three primary metrics—coefficient of determination ($R^2$), mean squared error (MSE), and root mean squared error (RMSE). The same metrics were also used to benchmark our results against three external datasets (ReGeom, GRDL, and in situ data from the Texas Water Development Board, TWDB). These metrics are calculated as follows:

$$R^2 = 1 - \frac{\sum_{i=1}^{n}(y_{iref} - y_{isim})^2}{\sum_{i=1}^{n}(y_{iref} - \overline{y_{ref}})^2} \quad \text{Equation 3}$$

$$MSE = \frac{\sum_{i=1}^{n}(y_{iref} - y_{isim})^2}{n} \quad \text{Equation 4}$$

$$RMSE = \sqrt{\frac{\sum_{i=1}^{n}(y_{iref} - y_{isim})^2}{n}} \quad \text{Equation 5}$$

where

- n is the number of data points,
- $y_{iref}$ is the $i^{th}$ reference value (from measurements or prior model outputs),
- $y_{isim}$ is the $i^{th}$ simulated value,
- $\overline{y_{ref}}$ is the mean of reference values.

A higher $R^2$ value signifies that the model effectively explains the variance in the data, while a lower RMSE indicates a closer alignment of the model's predictions with the observed values.

## 2.4 Data filtering and iterative outlier removal

We applied a systematic data-cleaning pipeline to filter out implausible values and extreme outliers that could bias the hydrological characteristics analysis. We began by merging data from HydroLAKES and GLOBathy, matching each record via the unique "Hylak_id" field so that the HydroLAKES' Shapefile attributes (e.g., average depth, residence time) could be paired with the corresponding GLOBathy's maximum depth. For every waterbody in this unified dataset, we then computed a depth ratio by dividing the average depth by the maximum depth and classified each entry as either a lake or a reservoir through HydroLAKES' variables. Next, we excluded all records with ratios ≤ 0 or >1, archiving them separately for potential future study. We also removed waterbodies displaying obviously incorrect residence times—specifically, values ≤ 0 or set to −9999, which "NA" values in original two databases. Many of these extreme residence times arose because the original discharge data used mean flow values, which were often very small, resulting in inflated residence time estimates. Finally, we applied an iterative outlier-removal procedure with a threshold of 1%, discarding only the single highest residence-time entry in each iteration if its removal caused more than a 1% shift in the mean. This process was then repeated until removing another outlier would alter the updated average by less than 1%, thus preventing a small number of extremely large values from disproportionately skewing the entire dataset.

## 3 Results and discussion

### 3.1 Hydrological characteristics analysis of global waterbodies

GLRDAV provides D-A-V equations and associated parameters for 1,427,688 waterbodies worldwide by combining HydroLAKES and GLOBathy records (Khazaei et al., 2022; Messager et al., 2016). Although many entries exhibit minor inconsistencies for the ratio $D_{ave}/D_{max}$ (e.g., $D_{ave}/D_{max} > 1$), these waterbodies were assigned the maximum depth from the GLOBathy database to ensure that all IDs receive at least one set of D-A-V relationships. In the analyses presented below, however, physically implausible and anomalous data were filtered out. Specifically, 44,588 waterbodies with $D_{ave}/D_{max} \leq 0$ or $D_{ave}/D_{max} > 1$ were flagged, along with additional exclusions based on invalid residence times, and a final iterative procedure removed the most extreme outliers in residence time. After filtering, 1,351,817 waterbodies remained (of which ~99.7% are lakes), forming the core dataset used in this chapter's hydrological assessments.

Globally (Figure 2), there are 1,351,817 lakes and reservoirs, covering a total surface area of $2.525 \times 10^6$ $km^2$ and a volume of $1.485 \times 10^5$ $km^3$, and an average hydraulic residence time of 1,783 days. For reservoirs alone, the average hydraulic residence time is 828 days. North America hosts 958,710 waterbodies, representing the largest count globally, with a total surface area of $1.2 \times 10^6$ $km^2$, a volume of $3.433 \times 10^4$ $km^3$, and an average residence time around 1,642 days (and 1,217 days for reservoirs). Europe, with 261,650 waterbodies, has the highest average residence time of 2,486 days and a total volume of $8.008 \times 10^4$ $km^3$, largely due to the Caspian Sea. Excluding the Caspian Sea reduces Europe's combined volume to $3.688 \times 10^3$ $km^3$. Areal, volume, and residence-time summaries for the other continents are provided in Supplementary Table S.1.

The geographical distribution of the water bodies included in GLRDAV is presented together with their corresponding $D_{ave}/D_{max}$ ratios in Figure 2. The $D_{ave}/D_{max}$ values reveal significant regional differences. Higher average ratios reflect flatter lake bottom morphologies, whereas lower ratios correspond to steeper shapes. The waterbodies in North America and Europe have similar mean $D_{ave}/D_{max}$ values of 0.34 and 0.38, respectively. Without the Caspian Sea, the $D_{ave}/D_{max}$ value for Europe remains 0.38. For Asia and Africa, the average ratios are 0.29 and 0.33. South America and Oceania have the lowest mean $D_{ave}/D_{max}$ ratios of 0.26 and 0.24, respectively, reflecting a greater contribution of relatively steeply sloped lakes.

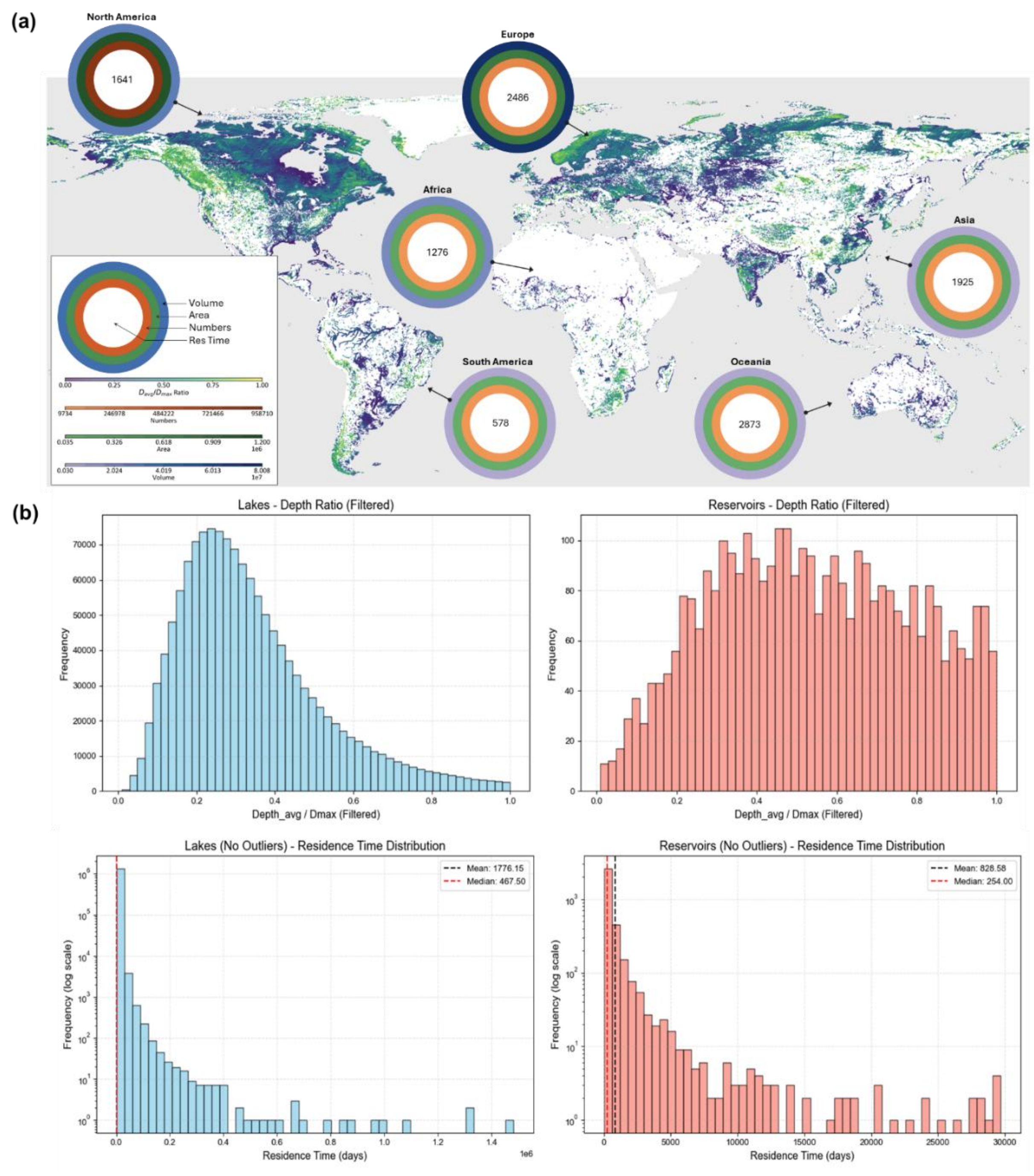

**Figure 2. (a) Global distribution of lakes and reservoirs. The circles present aggregated indicators for waterbodies by continent: the innermost white ring displays the average hydraulic residence time (days), while the rings denote the number, surface area ($km^2$), and volume ($km^3$) of waterbodies with darker colors correlating with higher values. The dots on the map indicate the geographical locations of the waterbodies while the color of the dots scales to the Dave / Dmax ratio of the water body. (b) Global distributions of the average-to-maximum depth ratio (top panel) and hydraulic residence time (bottom panel) for lakes (blue) and reservoirs (orange), following the removal of physically implausible values and extreme outliers.**

## 3.2 D-A-V relationship profile

Each water body in GRLDAV is represented by 12 equations that define its D-A-V characteristics, including six D-A equations and six D-V equations. Each set of six equations encompasses five polynomial functions, spanning first- to fifth-order terms, along with a power function equation. Figure 3 illustrates the D-A and D-V relationships for eight randomly selected waterbodies. Additionally, GLRDAV provides for each water body the coordinates (matching those of the ID in HydroLAKES and GLOBathy) and a CSV file with area and volume values for every 0.1-meter depth interval.

Compared to existing databases, GLRDAV offers more user-flexibility and greater depth resolution, while maintaining comparable accuracy. It encompasses smaller water bodies often omitted in earlier datasets, despite their importance in carbon cycling, nutrient dynamics, and climate regulation (Donchyts et al., 2022; Pi et al., 2022). The inclusion of both polynomial and power functions enables the user to choose the option that best fits their application. Although higher-order polynomials (order 4 and 5) generally deliver superior fits, particularly for large or morphologically complex waterbodies, they may yield negative slopes at the lower end of the water depth range, as well as negative A and V values. Alternatively, in certain cases, a combination of equations, say, using the power function at shallower water depths and a higher-order polynomial at greater depths, can produce accurate and physically valid A and V values across the entire water depth range.

When the fitted equations are evaluated against the original GLOBathy and HydroLAKES bathymetry, the mean $R^2$ exceeds 0.80 and the median RMSE for D-V curves is below $1 \times 10^{-4}$ $km^3$. Fourth and fifth-order polynomials give the tightest fits, but the power law performs nearly as well for small, shallow lakes. Full error distributions and example plots have been relocated to supplementary for reference.

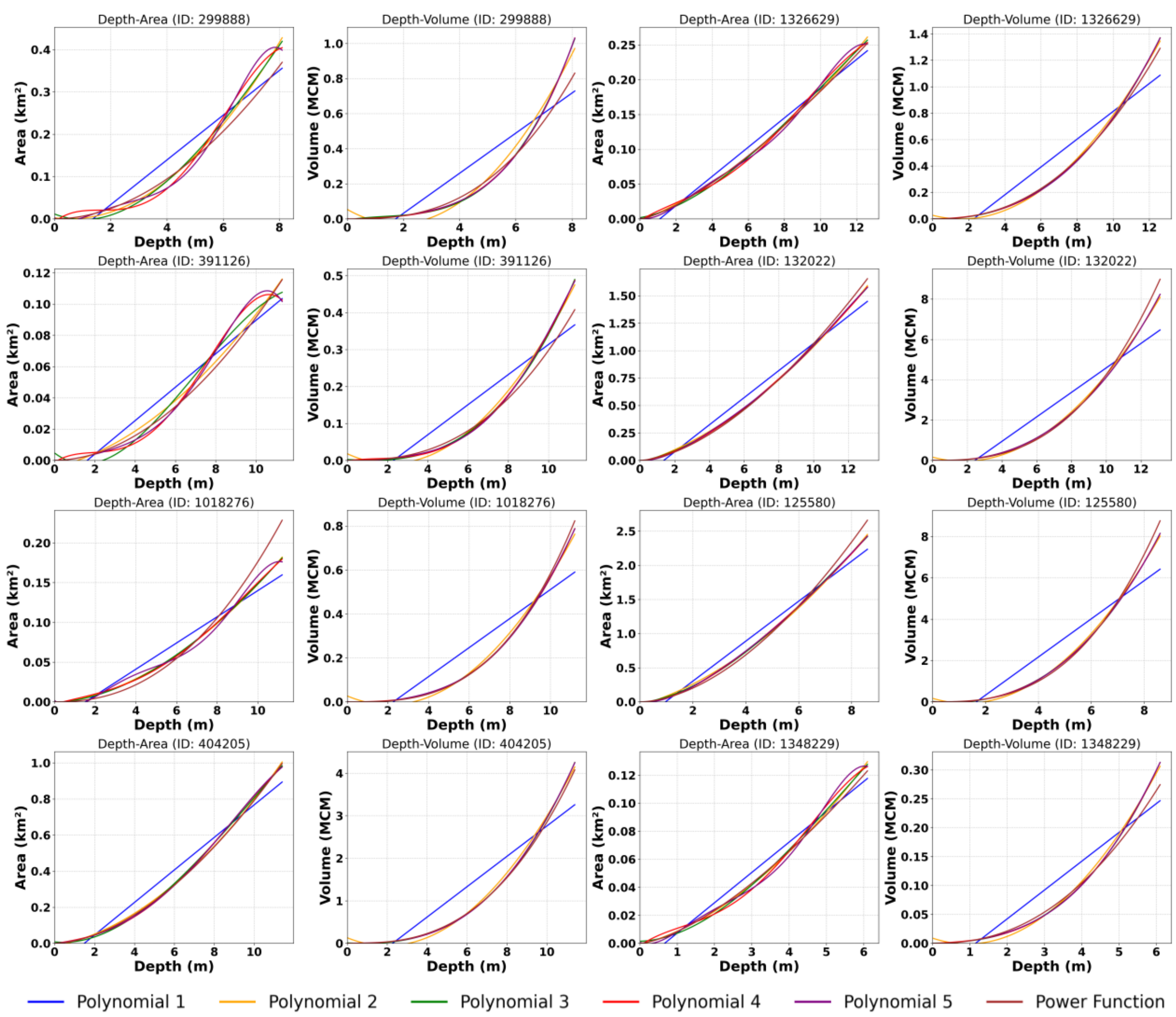

**Figure 3. D-A and D-V relationships for eight randomly selected waterbodies in GLRDAV. Each panel compares the results of polynomial regressions (orders 1 to 5) and a power function. Polynomial segments that yield negative area or volume are omitted; only physically meaningful depths (≥ 0 m) are shown.**

### 3.3 Comparison with ReGeom and GRDL relationships

We validated the GLRDAV-derived D-A and D-V relationships against two independent datasets—ReGeom (Yigzaw et al., 2018) and GRDL (Hao et al., 2024) using RMSE as the evaluation metric. Overall, the models show satisfactory performance, with average D-A RMSE ranging from 27.36 $km^2$ up to 35.60 $km^2$, and D-V RMSE values spanning 0.192 $km^3$ to 1.213 $km^3$, despite the diverse methodologies used to compile these external databases (Table 2).

**Table 2. Comparative comparison of GLRDAV polynomial (orders 1–5) and power function models against the ReGeom and GRDL datasets. RMSE values are reported for depth–area (D-A) and depth–volume (D-V) relationships, illustrating how each model performs across two independent global waterbody databases.**

| Metric | Polynomial Function Order 1 | Polynomial Function Order 2 | Polynomial Function Order 3 | Polynomial Function Order 4 | Polynomial Function Order 5 | Power Function |
|---|---|---|---|---|---|---|
| **Validation against ReGeom Dataset** | | | | | | |
| **RMSE (D-A)** | 27.26 | 27.87 | 27.38 | 27.65 | 27.60 | 27.29 |
| **RMSE (D-V)** | 0.650 | 0.305 | 0.193 | 0.192 | 0.192 | 0.192 |
| **Validation against GRDL Dataset** | | | | | | |
| **RMSE (D-A)** | 34.29 | 35.38 | 35.52 | 35.60 | 35.59 | 35.50 |
| **RMSE (D-V)** | 1.213 | 1.091 | 1.043 | 1.047 | 1.050 | 1.057 |

As summarized in Table 2, polynomial models of orders 3, 4, and 5 perform similarly across both ReGeom and GRDL datasets. In ReGeom, their D–A RMSE ranges from 27.38 to 27.65 $km^2$, and D–V RMSE from 0.192 to 0.65 $km^3$. In GRDL, the same models yield D–A RMSE between 35.52 and 35.60 $km^2$, and D–V RMSE from 1.043 to 1.050 $km^3$. In contrast, the 1st-order polynomial model yields slightly lower D-A RMSE values (27.26 $km^2$ for ReGeom and 34.29 $km^2$ for GRDL), but it is prone to producing negative predictions—especially in shallow or non-linear segments—due to its intercept term. This limitation underscores the necessity of supplementing low-order models with the power function, which, for ReGeom, achieves the lowest D-A (27.29 $km^2$) and D-V (0.192 $km^3$) RMSE. Nonetheless, the power function does not consistently outperform higher-order polynomials in all cases, suggesting that a combined approach is preferable.

Although GRDL employs extensive machine learning methods with significant computational resources, its RMSE values for D-A (35.5-35.6 $km^2$) and D-V (1.043–1.213 $km^3$) are comparable to those obtained by simpler polynomial approaches. This indicates that increased computational complexity does not automatically translate to better predictive performance, and it may impede usability in future research. Figures 4 and 5 illustrate the model performance for randomly selected waterbodies. In the D-A validation (Figure 4), higher-order polynomial functions (orders 3–5) provide a closer fit to both ReGeom (orange circles) and GRDL (black circles) outputs, whereas the 1st-order polynomial shows higher deviations—often yielding negative values in shallow areas. The power function, while less precise than the higher order models, remains a robust alternative for waterbodies with simpler bathymetries. In the D-V validation (Figure 5), polynomial order 5 consistently achieves the lowest RMSE, although the power function remains competitive in some cases. Figure 6 presents scatter plots of log (RMSE) values for D-A and D-V relationships across the whole ReGeom and GRDL waterbodies. The data points reveal that while the 5th-order polynomial generally attains lower errors overall, the power function has less complex D-A-V dynamics. By contrast, the 1st-order polynomial exhibits larger deviations, particularly for deeper or irregular waterbodies. Additionally, the upward trend in RMSE with increasing depth underscores the difficulty of capturing complex bathymetries using a single functional form, thereby reinforcing the importance of combining multiple model types to accommodate diverse underwater topographies.

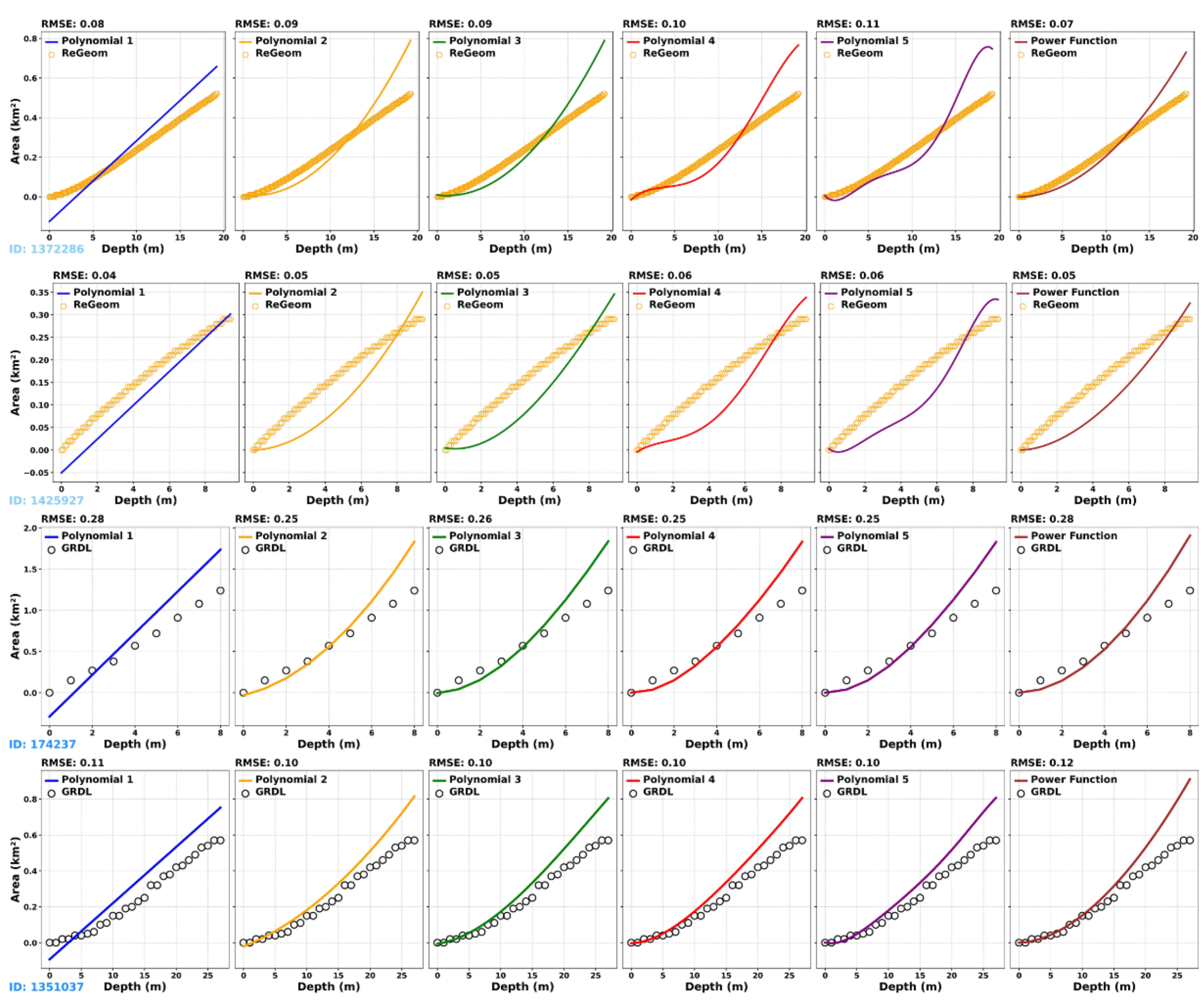

**Figure 4. Comparison of the depth–area (D-A) relationship against the ReGeom (orange circles) and GRDL (black circles) datasets for selected waterbodies (ID in blue text).**

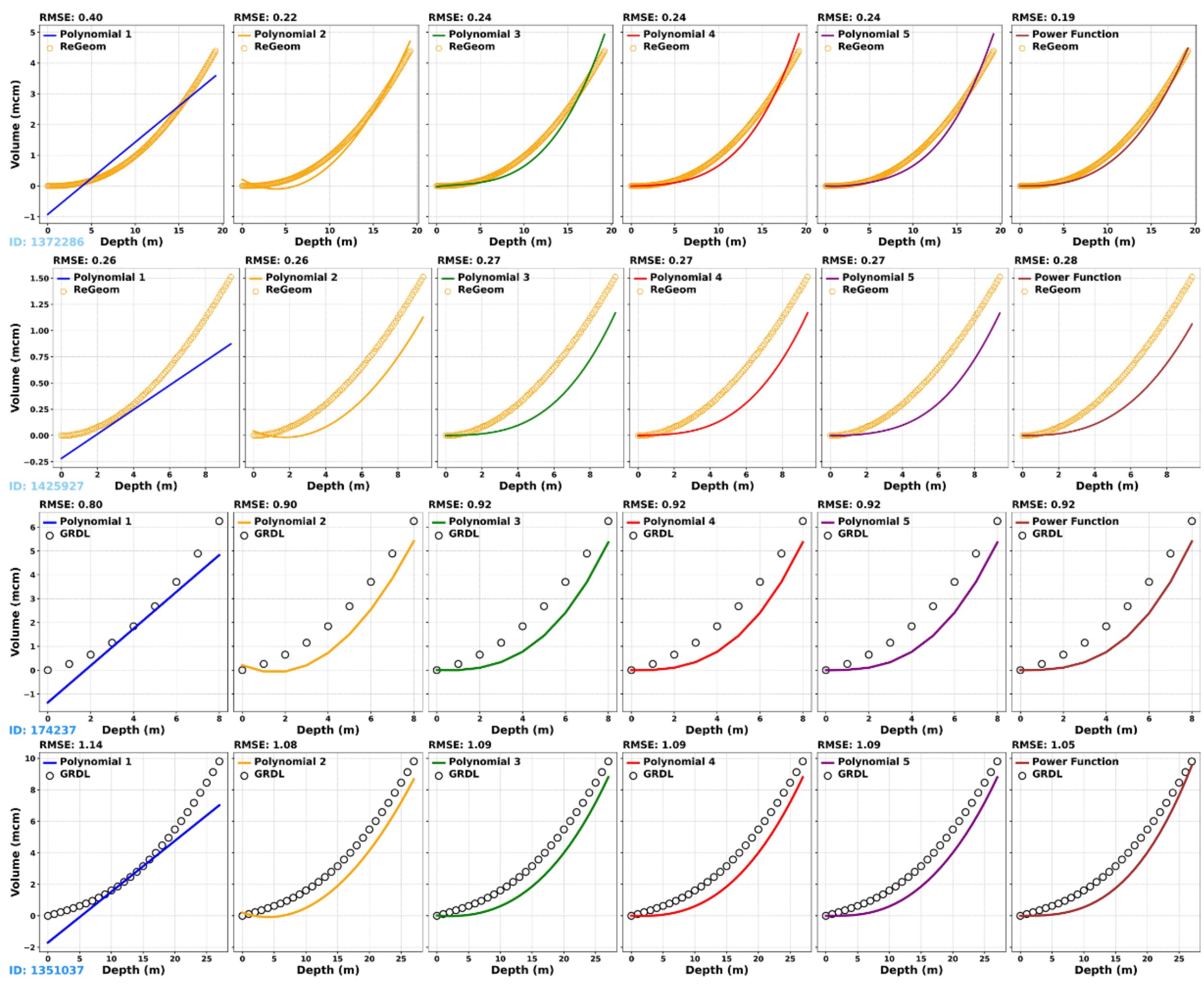

**Figure 5. Comparison of the depth–volume (D-V) relationship against the ReGeom (orange circles) and GRDL (black circles) datasets for selected waterbodies (ID in blue text).**

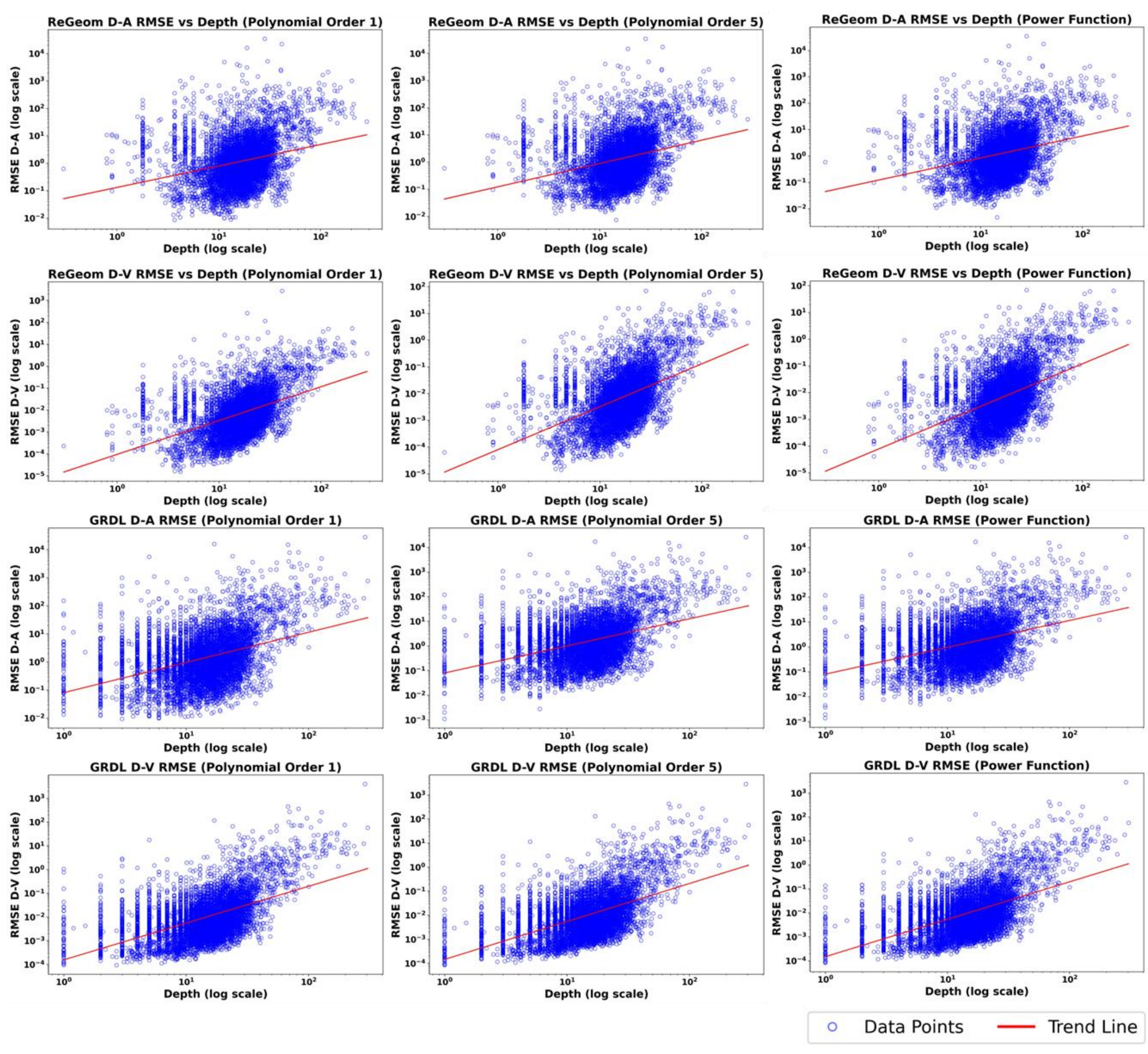

**Figure 6. Scatter plots of RMSE (log scale) values for D-A and D-V relationships of ReGeom and GRDL waterbodies, models of polynomial orders 1 and 5, and power function included.**

## 3.4 Comparison with in-situ data

To test the performance of the GLRDAV-derived D-A-V relationships, we compared model predictions with in-situ measurements from the Texas Water Development Board (TWDB; https://www.twdb.texas.gov/surfacewater/surveys/completed/list/index.asp). Four waterbodies—Cedar Creek, Coleman Reservoir, Hubbard Creek, and Lake Whitney—were selected for analysis because they are included in all datasets. For each waterbody, predictions from GLRDAV (using polynomial functions of orders 1–5 and a power function) were compared against results from ReGeom and GRDL (**Figure 7**).

**(1) Cedar Creek: for the D-A relationship, Table S.3 shows that the GRDL model yields an RMSE of 23.0 km² ($R^2$ = 0.73), while the ReGeom result is slightly improved (RMSE = 16.9 km2, R2 = 0.85). In contrast, the GLRDAV models based on polynomial functions of orders 2–5 achieve RMSE values of 11.9 km² with R2 values close to 0.93, representing an improvement of approximately 35% in RMSE compared to both GRDL and the power function (which has an RMSE of 18.3 km2 and R2 of 0.83). For the D-V relationship, GRDL performs very poorly (RMSE = 0.248 km3, R2 = –0.044), while ReGeom gives a considerably better result (RMSE = 0.065 km3, R2 = 0.93). Here, the GLRDAV higher-order polynomial models (orders 3–5) produce RMSE values of about 0.079 km3 with R2 near 0.90, whereas the power function's RMSE is 0.114 km3 (R2 = 0.78).**

**(2) Coleman Reservoir: our results indicates that for the D-A relationship, the GRDL model is highly inconsistent (RMSE = 2.3 km2 with a negative R²), while ReGeom achieves an RMSE of 0.65 km2 (R2 = 0.60). Notably, the GLRDAV 1st-order polynomial model delivers the best performance (RMSE = 0.4 km2, R2 = 0.83), with the power function (RMSE = 0.6 km2, R2 = 0.62) and other polynomial orders showing slightly higher errors. In the D-V relationship, GRDL again shows abnormal results (RMSE = 0.014 km3, R2 = –10.22), and ReGeom obtains an RMSE of 0.004 km3. Meanwhile, the GLRDAV polynomial models (orders 2–5) yield RMSE values in the narrow range of 0.0023–0.0024 km3 with R2 values around 0.68–0.69, outperforming both GRDL and the power function (RMSE = 0.0024 km3, R2 = 0.67) (Table S.4).**

**(3) Hubbard Creek: As shown in Table S.5, for the D-A relationship, the GLRDAV power function stands out with an RMSE of 1.9 km2 and an R2 of 0.95, whereas the polynomial models (orders 1–5) yield RMSE values around 5.5 km2 and R² values near 0.61. For the D-V relationship, although the higher-order polynomial models (orders 3–5) achieve RMSE values around 0.007 km3 with R2**

**near 0.96, the power function produces an even lower RMSE of 0.005 km3 and a higher R2 of 0.98, indicating very good performance in this case.**

**(4) Lake Whitney: Table S.6 provides that for the D-A relationship, the power function again performs best with an RMSE of 1.4 km2 and an R2 of 0.72, slightly outperforming the GRDL model (RMSE = 1.5 km2, R2 = 0.71) and the polynomial models (RMSE = 1.6 km2, R2 = 0.65). In contrast, for the D-V relationship, GRDL exhibits the lowest RMSE at 0.002 km3 with an $R^2$ of 0.96, while the power function achieves an RMSE of 0.003 km3 (R2 = 0.95) and the polynomial models (orders 2–5) have higher RMSE values around 0.004 km3 (R2 = 0.87).**

These comparisons demonstrate that the performance of the GLRDAV models varies with waterbody D-A-V relationship. In general, for waterbodies such as Cedar Creek and Coleman Reservoir, higher-order polynomial models (orders 3–5) consistently provide well performance, reducing RMSE by up to 35% compared to other models and outperforming ReGeom and GRDL. However, for Hubbard Creek and Lake Whitney, the power function occasionally delivers better results for one or both relationships. Overall, while the power function has been widely used in prior studies, our analysis shows that its performance is comparable only to that of the 2nd-order polynomial; in most cases, higher-order polynomial models yield greater accuracy and stability. Additionally, the external datasets GRDL and ReGeom exhibit notable inconsistencies, with GRDL showing extremely poor performance for certain waterbodies. Thus, the GLRDAV framework's flexibility—offering multiple functional forms—enables users to select the optimal model based on specific waterbody characteristics, such as water bodies' IDs and user defined depth.

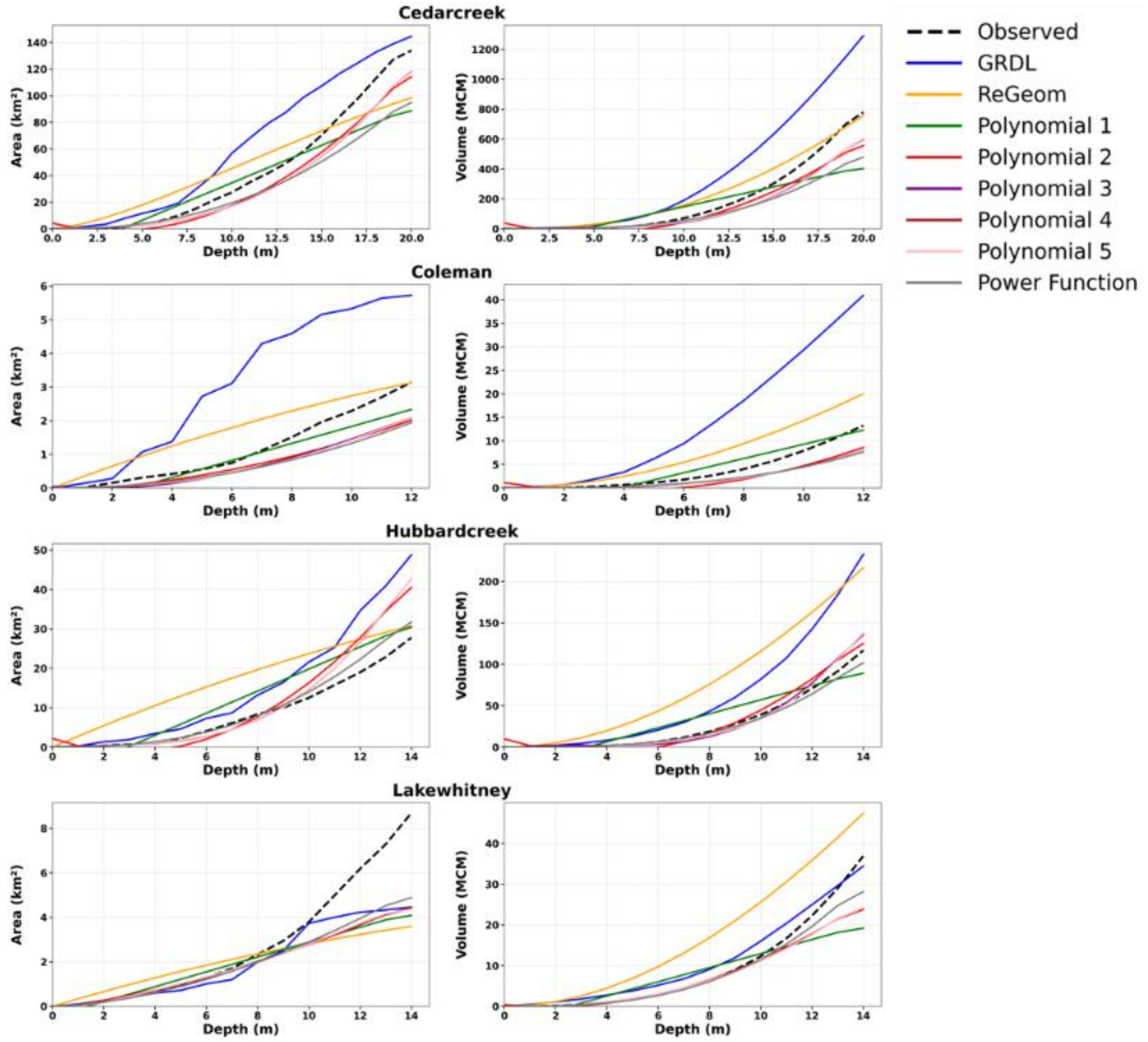

**Figure 7. Comparison of GLRDAV equations (polynomial orders 1–5 and power function) and ReGeom and GRDL models versus in-situ data-based D-A and D-V relationships for four waterbodies in Texas (Cedar Creek, Coleman Reservoir, Hubbard Creek, and Lake Whitney). In general, GRDL and ReGeom exhibit poorer fits to the observed relationships. Note that none of the model equations capture the pronounced change in slope of the observed D-A curve of Lake Whitney.**

## 3.5 Global Waterbody Calculator usage note and examples

To streamline the application of the GLRDAV equations in water quality modeling and future studies, we developed a Python package called "Global Waterbody Calculator" (Figure 8). This package, available via PyPI, integrates several libraries—SymPy for symbolic equation parsing, Geopy for distance-based queries, Pandas for data handling, and Matplotlib for visualization. Once installed, users can specify water bodies either by their unique IDs (consistent with HydroLAKES IDs) or by latitude and longitude coordinates. If only coordinates are provided, the tool automatically identifies the nearest water body using geospatial distance calculations.

After selecting the waterbodies of interest, the package computes D-A and D-V relationships using the polynomial (orders 1–5) and power function coefficients derived from GLRDAV. Results are presented as Pandas DataFrames and can be exported to CSV files for further processing. The package also generates optional plots showing how area and volume vary with depth, typically at 0.1 m increments. This functionality supports both single-waterbody analyses and batch processing for larger datasets, enabling efficient exploration of hydrological and ecological scenarios on regional or global scales.

To illustrate how the GLRDAV equations can be converted into spatially explicit bathymetry, we reconstructed 3-D maps for four water bodies—Lake Erie, Cedar Creek Reservoir, Lake Ontario, and Lake Diefenbaker—using the "Global Waterbody Calculator" (Figure 9). The resulting surfaces capture each basin's distinctive bathymetric signature—Erie's broad littoral shelves, the steep dendritic arms of Cedar Creek, Ontario's twin basins and the pool-riffle rhythm of Lake Diefenbaker—delivering the spatial fidelity required for process-based models, in which shoreline geometry shapes wave energy and resuspension while bottom slope governs deep-water circulation and nutrient exchange. By automating this reconstruction, the calculator removes a long-standing barrier to basin-scale assessments and opens the GLRDAV framework to wider applications, from mapping pollutants distributions and delineating critical habitats to tracking climate-driven shifts in water storage across thousands of lakes worldwide.

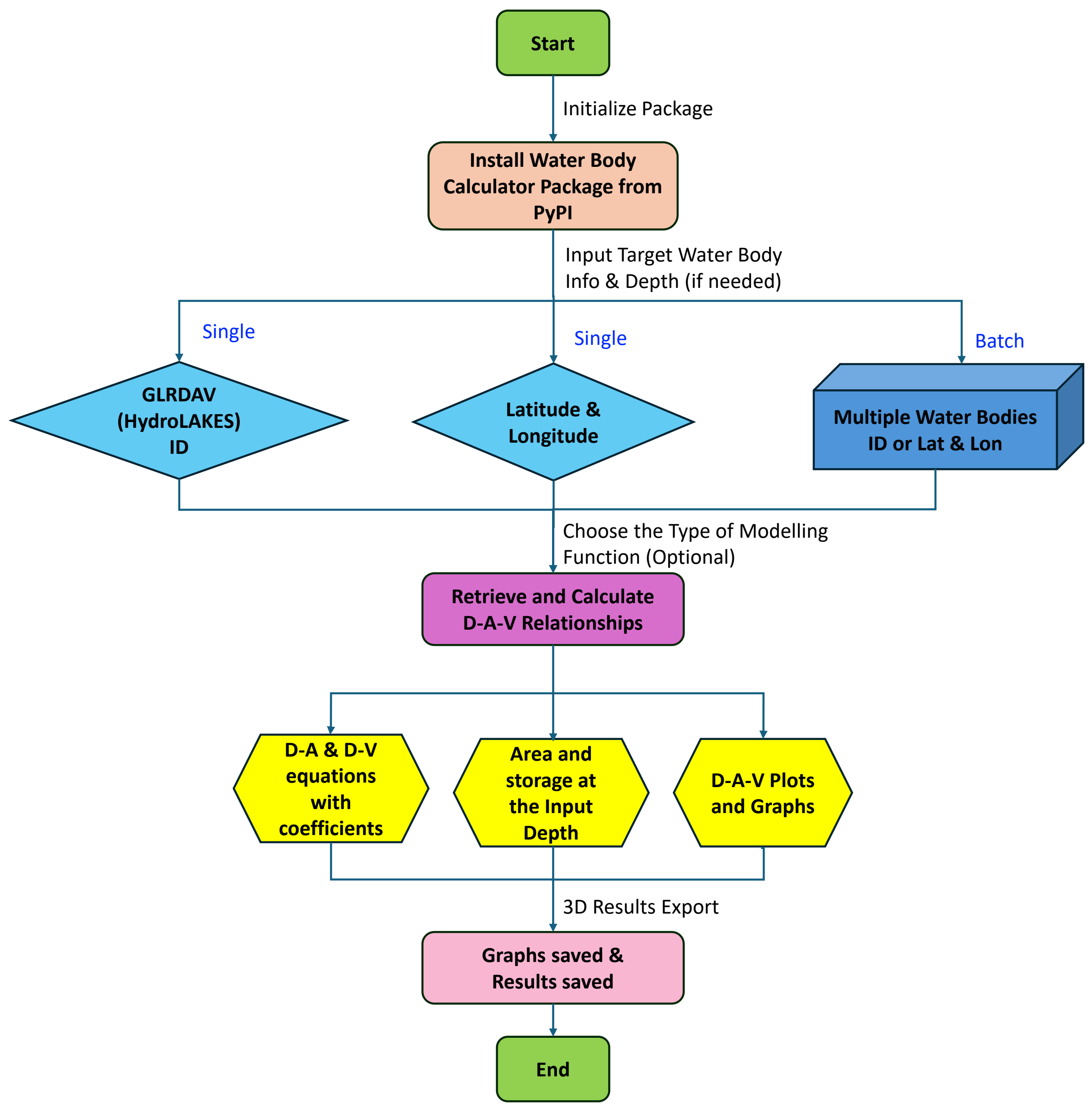

**Figure 8. Workflow of the "Global Waterbody Calculator" Python package. Users can specify a single water body (by ID or geographic coordinates) or process multiple water bodies in batch mode. The tool retrieves the relevant D-A-V equations from GLRDAV, calculates area and volume at specified depths, and outputs both tabular and graphical results for further analysis.**

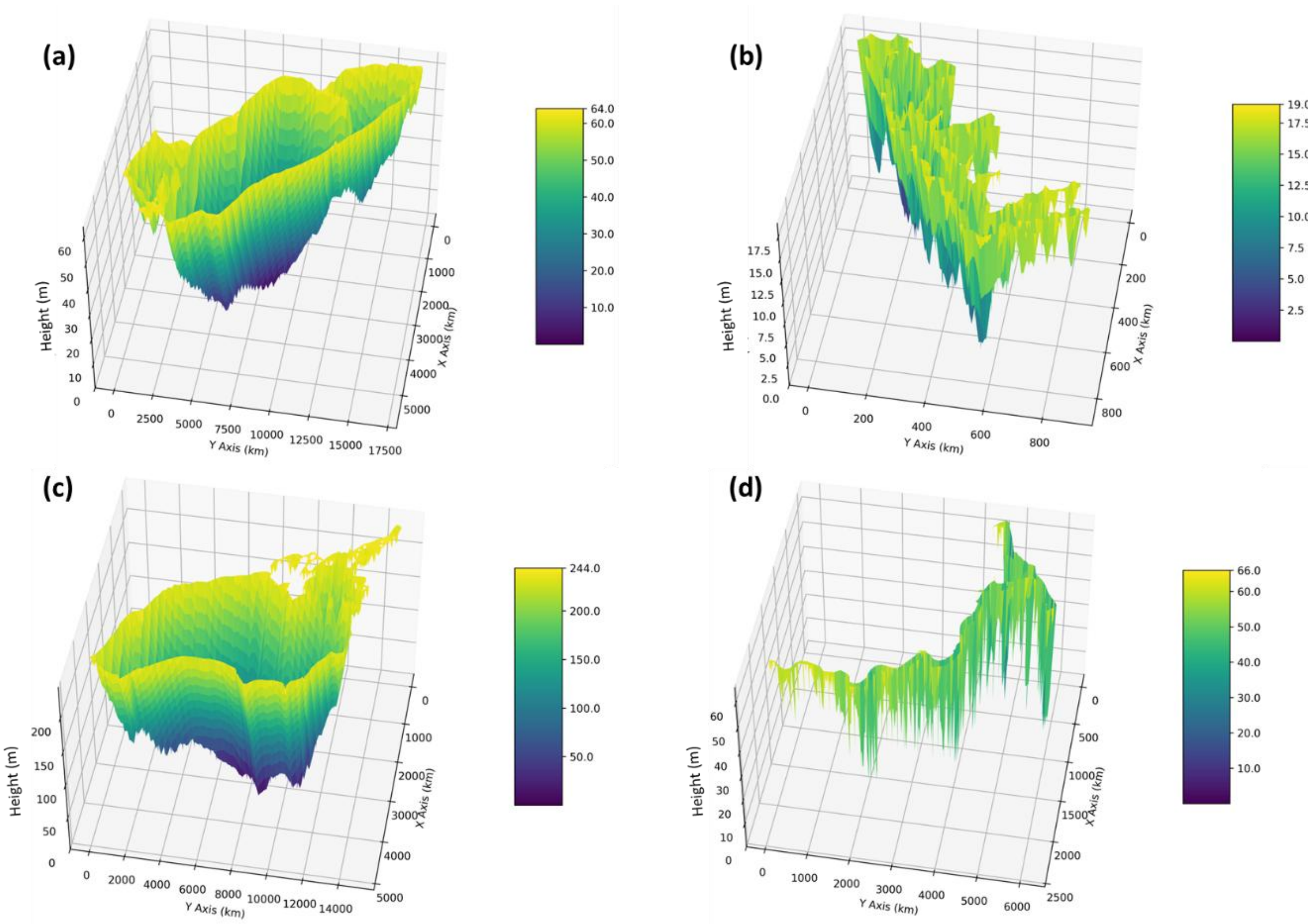

**Figure 9. Three-dimensional bathymetry generated with the Global Waterbody Calculator from the GLRDAV D-A-V equations for four water bodies: (a) Lake Erie (maximum depth ≈ 66 m), (b) Cedar Creek Reservoir (maximum depth ≈ 19 m), (c) Lake Ontario (maximum depth ≈ 244 m), (d) Lake Diefenbaker (maximum depth ≈ 66 m).**

## 4 Conclusion

The GLRDAV database addresses longstanding issues of the global waterbody D-A-V relationships, such as unit inconsistencies, oversimplified D-A-V equations, low spatial resolution, data errors, and cumbersome computational requirements. The GLRDAV database comprises over 17 million equations for over 1.4 million lakes and reservoirs with 0.1 m depth resolution, and 12 equations per waterbody (6 for depth–area and 6 for depth–volume relationships) presented as polynomial (orders 1–5) and power functions.

Against the ReGeom and GRDL datasets, 3rd–5th-order fits keep depth–volume RMSE near 0.20 km³ for ReGeom and ≈1.05 km³ for GRDL, about 10–20% lower than second-order or power-law forms while avoiding the negative predictions seen with first-order polynomials. Independent surveys from the Texas

Water Development Board show the same equations more closely track measured bathymetry than either GRDL or ReGeom.

Importantly, our results underscore that more sophisticated and computationally demanding machine learning models, such as those employed in GRDL, do not necessarily guarantee a better performance. Instead, our approach—leveraging existing large-scale datasets and combining polynomial and power function formulations—is both computationally efficient and of wide practical applicability.

GLRDAV is distributed with an open-source Global Waterbody Calculator that converts any record to tabular or GeoTIFF bathymetry within seconds, enabling rapid coupling to hydrodynamic or biogeochemical models such as CE-QUAL-W2, EFDC, and Delft3D. Together, the database and toolset provide the first globally consistent, high-resolution morphometric foundation for continental-scale assessments of water storage, mixing, greenhouse-gas emissions, and climate-driven change.

## 5 Data and tool availability

The GLRDAV database and associated depth–area–volume relationship datasets are publicly available through the Federated Research Data Repository (FRDR) at https://doi.org/10.20383/103.0881. The dataset, titled "Global lakes depth-area-volume relationship with 0.1 m resolution," provides depth–area and depth–volume relationships for more than 1.4 million global lakes and reservoirs at 0.1 m depth resolution. Python package is publicly available at https://pypi.org/project/globalwaterbodycalculator/

### Acknowledgements

The authors thank the Global Water Futures (GWF) program for supporting this research.

### Funding

This research was supported by Shengde Yu's research funding, the Global Water Futures (GWF) program funded by the Canada First Research Excellence Fund (CFREF), and the Natural Sciences and Engineering Research Council of Canada (NSERC).

## 6 CRediT authorship contribution statement

S.Y. conceived the study, designed the GLRDAV framework, developed the Global Waterbody Calculator, processed and integrated the global waterbody datasets, performed the depth–area–volume equation fitting and validation analyses, generated the figures, and wrote the original manuscript. Y.W. contributed to data processing, computational implementation, and Python package. W.L. contributed to data analysis, and Python package. P.V.C. supervised the study, provided conceptual guidance, contributed to the interpretation of the results, acquired funding, and reviewed and edited the manuscript. All authors discussed the results and approved the final version of the manuscript.

## 7 Competing Interests

The authors declare no competing interests.

# 9 Supplementary

## 9.1 Supplementary tables:

**Table S.1. Summary of global fresh waterbodies hydrological characteristics.**

| Continent | Type | Count | Total_Volume | Mean_Res_time | Mean_Ratio | Median_Ratio | Total_Lake_Area |
|---|---|---|---|---|---|---|---|
| **Global** | All | 1351817 | 1.53E+08 | 1783.417 | 0.341067 | 0.302771 | 2525163 |
| **Africa** | All | 14284 | 28736348 | 1276.054 | 0.326801 | 0.253842 | 189957.7 |
| **Asia** | All | 57864 | 5657873 | 1924.831 | 0.293006 | 0.224037 | 224695.7 |
| **Europe** | All | 261650 | 80075070 | 2486.03 | 0.376932 | 0.339363 | 743723.3 |
| **North America** | All | 958710 | 34325322 | 1641.933 | 0.339429 | 0.303157 | 1199598 |
| **Oceania** | All | 9734 | 299514.5 | 2873.061 | 0.237954 | 0.195657 | 35432.22 |
| **South America** | All | 49575 | 3705349 | 578.4185 | 0.263915 | 0.197873 | 131756.3 |
| **Europe** | Lake (NonCaspian) | 261138 | 3687591 | 2489.409 | 0.376533 | 0.339076 | 321793.2 |
| **Africa** | Lake | 13849 | 27991502 | 1297.432 | 0.322072 | 0.24857 | 158241.8 |
| **Africa** | Reservoir | 435 | 744846.1 | 595.4552 | 0.47738 | 0.454721 | 31715.87 |
| **Asia** | Lake | 56952 | 4797084 | 1947.806 | 0.288279 | 0.221013 | 194165.4 |
| **Asia** | Reservoir | 912 | 860789.6 | 490.0842 | 0.588196 | 0.590132 | 30530.23 |
| **Europe** | Lake | 261139 | 79287591 | 2489.813 | 0.376533 | 0.339076 | 698795.1 |
| **Europe** | Reservoir | 511 | 787478.3 | 553.0413 | 0.58114 | 0.565233 | 44928.26 |
| **North America** | Lake | 957361 | 33212295 | 1642.531 | 0.339185 | 0.303009 | 1129598 |
| **North America** | Reservoir | 1349 | 1113027 | 1217.117 | 0.513019 | 0.497832 | 70000.39 |
| **Oceania** | Lake | 9604 | 240926 | 2895.338 | 0.233398 | 0.193683 | 32302.73 |
| **Oceania** | Reservoir | 130 | 58588.5 | 1227.306 | 0.574531 | 0.600855 | 3129.49 |
| **South America** | Lake | 49400 | 2972885 | 578.0397 | 0.263089 | 0.197265 | 98534.39 |
| **South America** | Reservoir | 175 | 732464 | 685.3411 | 0.497187 | 0.463275 | 33221.91 |

**Table S.2. Comparative performance of D-A-V relationships: $R^2$ and RMSE compared with training dataset.**

| Metric | Polynomial Function Order 1 | Polynomial Function Order 2 | Polynomial Function Order 3 | Polynomial Function Order 4 | Polynomial Function Order 5 | Power Function |
|---|---|---|---|---|---|---|
| **$R^2$ (D-A)** | 0.9300 | 0.9815 | 0.9864 | 0.9908 | 0.9935 | 0.9612 |
| **$R^2$ (D-V)** | 0.8363 | 0.9943 | 0.9997 | 0.9999 | 0.9999 | 0.9810 |
| **RMSE (D-A)** | 0.22976 | 0.06145 | 0.03239 | 0.01955 | 0.01284 | 0.12072 |
| **RMSE (D-V)** | 0.02103 | 0.00435 | 0.00037 | 0.00015 | 0.00007 | 0.00483 |

**In situ data comparison summary (Area unit: $km^2$, RMSE unit: mcm):**

**Table S.3. Statistical analysis of Cedar creek in situ data with GLRDAV regression results.**

| Metric | Model | RMSE | $R^2$ |
|---|---|---|---|
| Depth-Area | GRDL | 23.01695 | 0.726266 |
| Depth-Area | ReGeom | 16.90552 | 0.85233 |
| Depth-Area | Polynomial 1 | 18.26107 | 0.8277 |
| Depth-Area | Polynomial 2 | 11.62189 | 0.930211 |
| Depth-Area | Polynomial 3 | 11.88291 | 0.927041 |
| Depth-Area | Polynomial 4 | 11.86833 | 0.92722 |
| Depth-Area | Polynomial 5 | 11.86535 | 0.927256 |
| Depth-Area | Power Function | 18.26808 | 0.827567 |
| Depth-Volume | GRDL | 247.5164 | -0.04385 |
| Depth-Volume | ReGeom | 64.8728 | 0.928294 |
| Depth-Volume | Polynomial 1 | 132.4282 | 0.701193 |
| Depth-Volume | Polynomial 2 | 80.19743 | 0.890415 |
| Depth-Volume | Polynomial 3 | 78.15776 | 0.895918 |
| Depth-Volume | Polynomial 4 | 78.53873 | 0.894901 |
| Depth-Volume | Polynomial 5 | 78.53701 | 0.894906 |
| Depth-Volume | Power Function | 113.5115 | 0.780462 |

**Table S.4. Statistical analysis of Coleman reservoir in situ data with GLRDAV regression results.**

| Metric | Model | RMSE | $R^2$ |
|---|---|---|---|
| Depth-Area | GRDL | 2.256451 | -3.78032 |
| Depth-Area | ReGeom | 0.653682 | 0.598821 |
| Depth-Area | Polynomial 1 | 0.422658 | 0.83228 |
| Depth-Area | Polynomial 2 | 0.572015 | 0.692801 |
| Depth-Area | Polynomial 3 | 0.571066 | 0.693819 |
| Depth-Area | Polynomial 4 | 0.583832 | 0.679977 |
| Depth-Area | Polynomial 5 | 0.582394 | 0.681551 |
| Depth-Area | Power Function | 0.636685 | 0.619412 |
| Depth-Volume | GRDL | 14.11172 | -10.2229 |
| Depth-Volume | ReGeom | 4.319339 | -0.05143 |
| Depth-Volume | Polynomial 1 | 2.524987 | 0.640694 |
| Depth-Volume | Polynomial 2 | 2.344741 | 0.690161 |
| Depth-Volume | Polynomial 3 | 2.340926 | 0.691168 |
| Depth-Volume | Polynomial 4 | 2.386441 | 0.679042 |
| Depth-Volume | Polynomial 5 | 2.38665 | 0.678986 |
| Depth-Volume | Power Function | 2.409027 | 0.672938 |

**Table S.5. Statistical analysis of Hubbard creek in situ data with GLRDAV regression results.**

| Metric | Model | RMSE | $R^2$ |
|---|---|---|---|
| Depth-Area | GRDL | 9.23523 | -0.11578 |
| Depth-Area | ReGeom | 8.757765 | -0.00339 |
| Depth-Area | Polynomial 1 | 5.377537 | 0.621688 |
| Depth-Area | Polynomial 2 | 5.479019 | 0.607275 |
| Depth-Area | Polynomial 3 | 5.488064 | 0.605978 |
| Depth-Area | Polynomial 4 | 5.486547 | 0.606195 |
| Depth-Area | Polynomial 5 | 5.506609 | 0.60331 |
| Depth-Area | Power Function | 1.889661 | 0.953286 |
| Depth-Volume | GRDL | 47.7362 | -0.74074 |
| Depth-Volume | ReGeom | 59.71937 | -1.72438 |
| Depth-Volume | Polynomial 1 | 16.90045 | 0.78181 |
| Depth-Volume | Polynomial 2 | 7.717923 | 0.954497 |
| Depth-Volume | Polynomial 3 | 7.202397 | 0.960373 |
| Depth-Volume | Polynomial 4 | 7.028161 | 0.962267 |
| Depth-Volume | Polynomial 5 | 7.00071 | 0.962561 |
| Depth-Volume | Power Function | 5.085508 | 0.980244 |

**Table S.6. Statistical analysis of Lake Whitney in situ data with GLRDAV regression results.**

| Metric | Model | RMSE | $R^2$ |
|---|---|---|---|
| Depth-Area | GRDL | 1.46278 | 0.712588 |
| Depth-Area | ReGeom | 1.937521 | 0.495758 |
| Depth-Area | Polynomial 1 | 1.716182 | 0.604385 |
| Depth-Area | Polynomial 2 | 1.612678 | 0.650665 |
| Depth-Area | Polynomial 3 | 1.615815 | 0.649305 |
| Depth-Area | Polynomial 4 | 1.615955 | 0.649245 |
| Depth-Area | Polynomial 5 | 1.609402 | 0.652083 |
| Depth-Area | Power Function | 1.432792 | 0.724252 |
| Depth-Volume | GRDL | 2.304644 | 0.958399 |
| Depth-Volume | ReGeom | 9.107509 | 0.350333 |
| Depth-Volume | Polynomial 1 | 6.012946 | 0.716817 |
| Depth-Volume | Polynomial 2 | 4.114506 | 0.867405 |
| Depth-Volume | Polynomial 3 | 4.049047 | 0.87159 |
| Depth-Volume | Polynomial 4 | 4.049074 | 0.871589 |
| Depth-Volume | Polynomial 5 | 4.051609 | 0.871428 |
| Depth-Volume | Power Function | 2.631349 | 0.945769 |

## 9.2 Readme and code:

**Global Waterbody Calculator**

Global Waterbody Calculator is an open-source Python package for estimating storage curves and bathymetry of freshwater bodies.

Given a HydroLAKES ID or geographic coordinates, the tool fits global depth-area-volume (D-A-V) relationships, exports results, and generates high-resolution GeoTIFFs plus interactive 3-D visualizations.

| Capability | Details |
|---|---|
| Global D-A & D-V retrieval | Based on HydroLAKES, GLOBathy, GLRDAV |
| 0.1 m resolution | Area & volume at 0.1 m depth steps |
| Publication-ready plots | CSV and PNG outputs |
| Bathymetric GeoTIFF | Raster created from lake polygons |
| Interactive 3-D view | Matplotlib/Plotly surface rendering |

**Features**

Global D-A and D-V retrieval based on HydroLAKES, GLOBathy, and GLRDAV relationships

0.1 m depth resolution for area and volume estimates

CSV and plot export for publication-ready outputs

Bathymetric GeoTIFF generation from lake polygons

Interactive 3-D visualization support

**Installation**

```
pip install globalwaterbodycalculator
```

**Requirements**

Python >= 3.7

Required packages installed via pip:

numpy

pandas

matplotlib

scipy

geopy

rasterio

scikit-learn

gdal

Ensure GDAL is installed and configured correctly on your system. It is required for raster and vector operations such as shapefile processing and TIFF output.

**Quick Start (Python API)**

```
from globalwaterbodycalculator.calculator import WaterBodyCalculator


# Initialize the calculator
calculator = WaterBodyCalculator()


# Calculate area-volume relationships by waterbody ID
result_df, water_body_id = calculator.calculate_area_volume(id=7,
depth=244)
calculator.save_results_to_csv(result_df, water_body_id, output_dir='.')
calculator.plot_results(result_df, water_body_id, output_dir='.')


# Alternatively, calculate using latitude and longitude
result_df, water_body_id =
calculator.calculate_area_volume(latitude=45.59193, longitude=47.71771,
depth=10)
calculator.save_results_to_csv(result_df, water_body_id, output_dir='.')
calculator.plot_results(result_df, water_body_id, output_dir='.')
```

**Bathymetric Mapping and 3-D Visualization**

You can generate bathymetric maps and 3-D plots from shapefiles:

```
calculator.generate_bathymetry_tiff(
    lake_id=7,          # Hylak_id of the waterbody
    shapefile='lakes.shp',  # Location of the shapefile
    id_field='Hylak_id',    # Name of the ID column
    depth=244,          # Maximum depth of the waterbody
    output_dir='output/',
    plot_3d=True        # Set to True to enable 3-D plotting
)
```

This will:

Compute the depth raster from the lake polygon and fitted D-A relationship

Save the bathymetric map as a GeoTIFF

Write an interactive 3-D surface plot of the lake basin

If plot_3d is False, only a GeoTIFF file will be generated

**Inputs**

id or latitude, longitude: HydroLAKES ID or WGS-84 coordinates

depth: Maximum depth (m)

Lake polygon (optional): ESRI Shapefile with an id_field matching Hylak_id

**Outputs**

<id>_dav.csv: Depth, area ($m^2$), volume ($m^3$)

<id>_dav.png: Area and volume curves

<id>_bathy.tif: Bathymetric raster

<id>_3d.html: Interactive 3-D view (optional)

**Directory Layout**

```
globalwaterbodycalculator/
├─ calculator.py       # Python API
├─ cli.py              # command-line entry point
├─ data_manager.py     # equation CSV download/cache helper
└─ __init__.py
```

**HydroLAKES Download and Licence**

Global Waterbody Calculator does not redistribute the HydroLAKES shapefiles. Bathymetric mapping and 3-D outputs require a HydroLAKES shapefile source.

To run volume and area calculations or create maps, you must:

Download the dataset from https://www.hydrosheds.org/products/hydrolakes.

Unzip it into a local folder, such as ~/data/HydroLAKES/.

Point the HYDROLAKES_DIR environment variable to that path or pass the folder to the API or CLI option --hydrolakes_path.